\documentstyle[12pt]{article}
\begin{document}
\thispagestyle{empty}
\begin{center} \LARGE \tt \bf {Slow decay of cosmic magnetic fields superadiabatically on curvature-torsion scales}
\end{center}
\vspace{1.0cm}
\begin{center}
{\large By L.C. Garcia de Andrade\footnote{Departamento de
F\'{\i}sica Te\'{o}rica - IF - UERJ - Rua S\~{a}o Francisco Xavier
524, Rio de Janeiro, RJ, Maracan\~{a}, CEP:20550.
e-mail:garcia@dft.if.uerj.br}}
\end{center}
\begin{abstract}
Recently Barrow, and Tsagas  [Phys Rev D 77: 107302, (2008)] have
shown that slow decay of cosmological magnetic slowly decay in FRW
universes on curvature scales as $B\sim{a^{-1}}$ in the context of
general relativity (GR). This helps possible amplification of cosmic
magnetic fields. In this paper starting from dynamo equations in
spacetimes with torsion we obtain also slow decay of magnetic fields
naturally on curvature-torsion scales of Riemann-Cartan spacetime on
a de Sitter universe. In this case the constant of proportionality
between the magnetic field and the curvature scale is the torsion in
the present universe. Thus the B-field becomes
$B\sim{\frac{{\eta}}{H_{0}}Ta^{-1}}$ where T is torsion vector
modulus, $H_{0}$ is the Hubble constant and $\eta$ is the diffusive
scale. The torsion effect is therefore enhanced by diffusive
processes in the universe. Galactic dynamo seeds of the order of
$10^{-26}Gauss$ are obtained. If one still uses the constant
torsion, and the scale of $10Kpc$ the galactic dynamo seed is given
by $B\sim{10^{-13}Gauss}$ which is lies very well within the
interval obtained for the cosmic magnetic field obtained by K.
Pandey et al [ApJ (2013)] for primordial magnetic field lines using
$Ly{\alpha}$ clouds on $1 Kpc$ scales, which is
$10^{-20}Gauss\le{B}\le{10^{-9}Gauss}$. In comoving cosmic plasmas
endowed with torsion cosmic magnetic fields as low as $10^{-25} G$
are obtained, which are still strong enough to seed galactic
dynamos.
\end{abstract}

Key-words: Torsion theories, primordial magnetic fields, galactic
dynamo seeds.
\newpage
\section{Introduction}

Earlier Barrow and Tsagas \cite{1} have shown that in the context of
GR, FRW universes magnetic field can experience superadiabatic
growth with B-field of the order of $B\sim{a^{-1}}$ instead of
$B\sim{a^{-2}}$ of the adiabatic growth. In their case the Maxwell
equations is used instead of more general Maxwell s equations used
to find galactic dynamo seeds \cite{2}. Coupling of vector fields to
spacetime geometry slows down the decay of large scale magnetic
field in open universe, compared to that seen in perfectly flat
models. This results in a large gain in B-field strength during the
pre-galactic era that leads to astrophysically interesting B-fields.
Seeds around $10^{-34}Gauss$ can sustain galactic dynamos if the FRW
universe is dark-energy dominated. It is well-known that dynamo
mechanism requires magnetic seeds with a collapse coherence length
of $100 pc$. This corresponds to a comoving (pre-collapse) scale of
approximately $10 Kpc$. Here we consider seed fields on
curvature-torsion scales of $10 Kpc$, which from dynamo equation
with torsion \cite{3}, yields a cosmological magnetic field of
$10^{-26}Gauss$ in contrast to the $10^{-22}Gauss$ obtained from
cosmic strings \cite{4}. If one still uses the constant torsion, and
the scale of $10Kpc$ the galactic dynamo seed is given by
$B\sim{10^{-10}Gauss}$ which is only one order of magnitude less
than the one obtained for the low limit of cosmic magnetic field
obtained by K. Pandey et al \cite{5} for primordial magnetic field
lines using $Ly{\alpha}$ clouds on $1 Kpc$ scales. To further test
the model we compute the cosmic magnetic fields in comoving cosmic
plasmas with torsion and obtain primordial magnetic seeds of the
order of $B\sim{10^{-25}Gauss}$ which though is weak is still enough
to seed galactic dynamos \cite{6}.
\section{Slow decay of cosmic magnetic fields with
torsion} It is important to say here that since torsion effects on
magnetic fields are enhanced by diffusion according to previous
obtained dynamo equation \cite{3}
\begin{equation}
{\partial}_{t}\textbf{B}-{\nabla}{\times}[\textbf{v}\times\textbf{B}]-{\eta}[{\Delta}+div\textbf{T}]
\textbf{B}=0 \label{1}
\end{equation}
we call the attention that the galactic magnetic fields are more
vulnerable to dissipation in the absence of a regenerated dynamo.
Dissipation of magnetic field is also associated with the presence
of magnetic monopoles \cite{7}. can be noticed from equation
(ref{1}) the fourth term on the left where divergence of torsion is
placed vanishes in highly conducting phases of the universe where
dissipation $\eta$ may be neglected. Let us now investigate based on
the dynamo equation above, the issue of the decay of magnetic field
on curvature-torsion scales. One shall prove that for a FRW universe
\begin{equation}
ds^{2}=dt^{2}-a^{2}d{\textbf{x}}^{2}\label{2}
\end{equation}
torsion scales do not change the carachter of the superadiabatic
amplification of the magnetic field but only adds a proportionality
factor linear on torsion and diffusion coefficient. By making use of
the partial differential operator
\begin{equation}
{\partial}_{t}=\frac{{\partial}a}{{\partial}t}\frac{{\partial}}{{\partial}a}\label{3}
\end{equation}
and substitution on dynamo equation yields
\begin{equation}
H_{0}{\partial}_{a}{B}={\eta}[\frac{1}{a^{3}}+\frac{T}{a^{2}}]B
\label{4}
\end{equation}
where use has been made of the Hubble parameter
$H_{0}=\frac{\dot{a}}{a}$ in this equation. Now the new dynamo
equation is simply solved to yield
\begin{equation}
B\sim{\frac{{\eta}}{H_{0}}Ta^{-1}} \label{5}
\end{equation}
which shows that for constant torsion the superadiabatically
expansion of the universe has a slower decay of the magnetic field
which is enhanced by torsion scales. Since
${\eta}\sim{10^{26}cm^{2}s^{-1}}$ and $a\sim{10Kpc}$,
$H_{0}\sim{5.5\pm{0.5}cm.sec^{-1}pc^{-1}}$ and
$T\sim{10^{-17}cm^{-1}}$ for torsion \cite{8}. Substitution of these
values into expression yields $B_{torsion}\sim{10^{-13}Gauss}$ which
fits very well within the limits of the interval obtained for the
cosmic magnetic field obtained by K. Pandey et al [ApJ (2013)] for
primordial magnetic field lines using $Ly{\alpha}$ clouds on $1 Kpc$
scales, which is $10^{-20}Gauss\le{B}\le{10^{-9}Gauss}$. Besides
this torsion scale bound obtained for the cosmic magnetic field is
strong enough to seed galactic dynamos. Actually it is also another
interval obtained by Barrow and Tsagas recently \cite{8} using FRW
GR metric.
\section{Magnetic fields in comoving plasmas with torsion}
In this section we shall show that the comoving cosmic plasmas
$(v\sim{0})$ with torsion induces a seed primordial field which
although is very weak is strong enough to seed galactic dynamos. Let
us now consider the dynamo equation using the following ansatz for
the cosmic magnetic field $B\sim{e^{{\gamma}t}}$. Substitution of
this ansatz in the dynamo equation in torsion scales above one
obtains
\begin{equation}
{\gamma}=-{\eta}[L^{-2}+TL^{-1}] \label{6}
\end{equation}
according to the above data and new data $L\sim{10^{28}cm}$ for the
radius of the universe, yields
\begin{equation}
|{\gamma}_{torsion}|\sim{10^{-19}s^{-1}} \label{7}
\end{equation}
Thus the B-field is given by
\begin{equation}
B\sim{B_{seed}{\gamma}_{torsion}{\Delta}t}
\end{equation}
which yields $B_{torsion}\sim{10^{-21}Gauss}$ for a seed field of
$10^{-20}Gauss$ such as in the lower bound obtained in (\ref{7}).
Thus this value is still a well within limits able to seed galactic
dynamos. For scale of $10Kpc$ one obtains
${\gamma}\sim{10^{-13}s^{-1}}$ which yields
$B_{torsion}\sim{10^{-25}Gauss}$ which is still a strong value to
seed galactic dynamos.
\section{Conclusions}
Therefore one may conclude that though five orders of magnitude
lower than the lower bound obtained astronomically this is still a
value able to seed galactic dynamos. We also know than the values of
the amplification factor ${\gamma}_{torsion}$ is very weak compared
to the values for fast dynamos, as ${\gamma}\sim{10^{8}s^{-1}}$
which leads us to conclude that the dynamos associated with torsion
are very slow. We hope to left more clear in this Brief Report the
role of torsion modes in the galactic dynamo seeds.

\section{Acknowledgements} We
would like to express my gratitude to D Sokoloff and A Brandenburg
and Prof C Sivaram for helpful discussions on the problem of dynamos
and torsion. Financial support from CNPq. and University of State of
Rio de Janeiro (UERJ) are grateful acknowledged.

\end{document}